\documentclass[aps, twocolumn, showpacs]{revtex4-1}
\usepackage{amsmath}
\begin{document}
\title{Topological approach to proton spin problem: decomposition controversy and beyond}
\author{S. C. Tiwari \\
Department of Physics, Banaras Hindu University, Varanasi 221005, \\ Institute of Natural Philosophy \\
Varanasi India\\}
\begin{abstract}
Lorentz covariant and gauge invariant definitions of quark and gluon spin and orbital angular momenta continue to pose a great theoretical challenge. A major controversy on the fundamental concepts
followed Chen et al proposal: the basic idea is to split the gauge potential into pure gauge and physical components motivated by the gauge symmetry. We term it gauge symmetry paradigm (GSP) to distinguish it
from the well-known inertial frame dependent transverse-longitudinal decomposition (TLP). A thorough study adhering to the traditional meaning of Lorentz covariance and gauge invariance is reported; it leads to a 
new result: logically consistent development of GSP does not exist and Chen et al proposal turns out to be either trivial or metamorphosed into TLP. Going beyond the controversy and the spin sum rules the necessity
for a nonperurbative QCD approach to address the proton spin problem is underlined. We suggest topological approach: generalized de Rham theorems for QCD, and spin as a topological invariant for baryons are discussed.
Nonabelian Stokes theorem is applied to derive color flux for the closed loop in a variant of Burkardt's U-shaped path. Similarity between Chen et al decomposition and Kondo decomposition of the gauge potential
is suggestive of a topological perspective on the Chen et al proposal with interesting physics.
 
\end{abstract}
\maketitle

\section{\bf Introduction}
In the nonrelativistic naive quark model of the hadrons the spin of the proton is attributed entirely to the spin of the constituent quarks. Beginning with the polarized deep inelastic scattering (pDIS) muon-proton experiments \cite{1}
many experiments over past twenty five years have established that quark spin content, though nonzero, is very small \cite{2}. Why? What is the origin of the proton spin? This is the heart of the proton spin problem that remains
unsolved. Quantum chromodynamics(QCD) is believed to be the dynamical theory for the quark-quark interaction, therefore it is natural to include the contribution of gluon angular momentum in the proton spin sum rule
$\frac{1}{2}=J^q+J^g$, where $J^q$ and $J^g$ denote the total angular momentum of quarks and gluons respectively; it is understood that these are the matrix elements of the corresponding quantum operators. In any free field or
interacting gauge theory it is well known that physical observables have to be gauge invariant. If we examine the spin sum rules then two important tasks emerge: separation of the total angular momentum into spin and orbital parts of
both quarks and gluons, and to define experimentally measurable quantities corresponding to them. Essentially it is the scattering cross section that one measures, and the data is used to extract information on the structure functions.
Quark-parton model or perturbative QCD calculations depend on the schemes and models \cite{2,3}. The pDIS experiments are performed with a preferred proton momentum direction, hence empirically, reasonable assumptions dispensing with
the strict adherence to the simultaneous manifest Lorentz covariance and gauge invariance could be made. Mainly due to this Jaffe-Manohar sum rule \cite{4} and Ji decomposition \cite{5} have proved to be of great utility \cite{2, 6}.

A new dimension to the proton spin problem was added with Chen et al proposal \cite{7}: "proton spin decomposition controversy" recently reviewed by Wakamatsu \cite{8} and Leader and Lorce \cite{9}. The main question is whether the
gauge invariant separation of spin and orbital angular momenta of quarks and gluons respecting relativistic invariance was indeed possible contrary to the textbook wisdom. Intense debate followed, and now seems to have been settled \cite{10,
11, 12} with the consensus that  manifest Lorentz covariance and gauge invariance cannot be satisfied simultaneously, see the earlier comment  \cite{13}, there are only two distinct angular momenta - kinetic and canonical, and experimental conditions
determine as to what is physically meaningful. One of the important fall outs of this debate is the reappraisal of conceptual issues and their pedagogical elucidation.

The convergence of the views seems superficial as the careful study of the literature \cite{6,10,11,12,14,15} shows, in particular the most glaring aspect is on the medley of gauge invariance and Lorentz invariance. Does Chen et al
proposal \cite{7} contain new physics? Is it in any way superior to Jaffe-Manohar and Ji decompositions?  Numerous papers on the controversy address the second question, however the first question remains untouched in the literature.
The aim of the present paper is to address these questions in the light of the recent papers on this subject, and to argue the logical conclusion of the critique that proton spin
problem has to be looked afresh going beyond the controversy and the conventional sum rules. A critique on fundamental questions, and new insights constitute the next section. The second question is addressed in Section III. 
Chen et al proposal is discussed in the context of the recent works of Wakamatsu, Lorce and Leader. It is suggested that the principal idea to split the gauge vector potential into pure and physical parts is based on gauge 
symmetry paradigm, while the usual longitudinal-transverse decomposition has frame of reference paradigm: if Chen et al proposition is analyzed in the strict gauge symmetry paradigm it is either trivial or turns out effectively
vacuous. Section IV presents topological approach \cite{16} to address the proton spin problem.

\section{\bf A critique and new insights}

A vast literature has come up on the controversy over the definitions and the decompositions of the total angular momentum into spin and orbital components of quarks and gluons since the publication of a provocative paper in 2008
\cite{7}. Two recent reviews devoted to the controversy \cite{8,9} indicate the importance of these questions for QCD physics. It seems the nature of the controversy, and the role of relativity and gauge symmetry have differing
perceptions \cite{17}. In spite of agreement on  some of the controversial points \cite{6,10,11,12,14,15} there do exist important questions lacking clarity and satisfactory resolution. A focused discussion on key questions to
throw new light on them is presented in this section.

${\bf Kinetic ~versus~ Canonical}$

In one of the earliest critique \cite{18} of \cite{7} Leader argues that canonical momentum and canonical angular momentum are the correct ones, and also presents a lucid discussion on some salient features of quantum electrodynamics
(QED) and QCD. In a recent contribution \cite{10} he explains the difference between kinetic and canonical versions. For the sake of clarity and completeness certain points have to be added on this issue. 
The formalism: Newtonian, Lagrangian or Hamiltonian, and
the nature of the dynamical system: classical, relativistic or quantum determine the choice of the dynamical variables. Physical interpretation, in most cases, is quite delicate. In the following we discuss few examples.

Podolsky \cite{19} considers a simple problem defined by a Lagrangian
\begin{equation}
L=\frac{1}{2}(m_1 v_1^2 +m_2 v_2^2) -U(R) + k {\bf v}_1.{\bf v}_2
\end{equation}
The kinetic or Newtonian momenta are $m_1{\bf v}_1$ and $m_2 {\bf v}_2$, but the angular momentum $ {\bf r}_1 \times m_1{\bf v}_1+{\bf r}_2 \times m_2{\bf v}_2$ is not conserved. Defining the canonical momenta $m_1 {\bf v}_1 +k {\bf
v}_2$ and $m_2 {\bf v}_2 +k {\bf v}_1$ the corresponding canonical angular momentum is conserved. Physical interpretation becomes somewhat delicate in the problem of electron-magnetic field and electric-magnetic
charge interactions. For an electron in a magnetic field canonical momentum $m{\bf v} -e{\bf A}$ gives the canonical angular momentum
\begin{equation}
{\bf L}_e ={\bf r} \times m{\bf v} -e{\bf r} \times {\bf A}
\end{equation}
For a particle with electric charge $e_1$ and magnetic charge $g_1$ in the field of a particle with charges $(e_2,g_2)$ the conserved angular momentum vector is found to be
\begin{equation}
{\bf J}_{eg} ={\bf r} \times m{\bf v} -(e_1g_2-e_2g_1) \frac{\bf r}{r}
\end{equation}
Schwinger \cite{20} explains that the second term in (3) could be interpreted as spin angular momentum, i. e. $-(e_1g_2-e_2g_1) =\frac{{\bf S}.{\bf r}}{r}$ ; then the angular momentum in (3) is the sum of orbital and spin components.
Note that Eq.(2) is not invariant under the gauge transformation
\begin{equation}
{\bf A} ~ \rightarrow ~ {\bf A} + {\bf \nabla} \chi
\end{equation}
However the gauge transformations have no role in the Newtonian approach: Schwinger follows this to calculate the rate of change of momentum equal to the Lorentz force, and the rate of change of angular momentum is given by 
the torque; thus the moment of momentum directly gives expression (3).

Interestingly the Hamilton-Jacobi equation has manifest gauge invariance
\begin{equation}
{({\bf \nabla}S -e{\bf A})}^2 -{(\frac{\partial S}{\partial t} +e \phi)}^2 =m^2
\end{equation}
The gauge transformation (4) along with the transformation
\begin{equation}
{\bf \nabla}S ~ \rightarrow ~ {\bf \nabla }S+e {\bf \nabla} \chi
\end{equation}
leaves the Hamilton-Jacobi equation invariant.

It may be mentioned that canonical momentum as a generator of the magnetic translation group is gauge covariant similar to ${\bf D} = {\bf \nabla} -ie{\bf A}$, and for the electron-monopole system the angular momentum is a gauge
invariant generator of rotations \cite{21}.

In field theory, it is well known that divergenceless canonical energy-momentum tensor, $T^{\mu\nu}$ is not gauge invariant and not symmetric for free EM field. In order to construct angular momentum tensor usually one symmetrizes
it. Here we give a short discussion on a nice approach following \cite{22}. Invoking invariance of the action under infinitesimal Lorentz transformation of the EM potentials one gets the spin density tensor
\begin{equation}
S_{\mu\nu\sigma} = F_{\mu\sigma} A_\nu -F_{\nu\sigma} A_\mu
\end{equation}
Here $F_{\mu\nu} = \partial_\mu A_\nu -\partial_\nu A_\mu$ is the EM field tensor, and $A_\mu$ is 4-vector potential. Spin energy-momentum tensor calculated from (7) added to the canonical tensor $T^{\mu\nu}$ gives the symmetric and
gauge invariant energy-momentum tensor
\begin{equation}
T^{\mu\nu}_s = - F^{\mu\lambda} F^\nu_\lambda +\frac{1}{4} g^{\mu\nu} F^{\alpha\beta} F_{\alpha\beta}
\end{equation}
Formally one can define angular momentum tensor using $T^{\mu\nu}$ to be
\begin{equation}
M_{\lambda\mu\nu} = x_\lambda T_{\mu\nu} - x_\mu T_{\lambda\nu}
\end{equation}
This tensor is not divergenceless, however the sum of (7) and (9) together with the addition of a divergenceless term gives
\begin{equation}
J_{\lambda\mu\nu} = S_{\lambda\mu\nu} +M_{\lambda\mu\nu} + \partial^\sigma (x_\lambda A_\mu F_{\nu\sigma} -x_\mu A_\lambda F_{\nu\sigma})
\end{equation}
After substituting various terms in (10) one gets the form  $x_\lambda F_{\nu\sigma} F^\sigma_\mu -x_\mu F_{\nu\sigma} F^\sigma_\lambda +\frac{1}{4} (x_\lambda g_{\mu\nu} -x_\mu g_{\lambda\nu}) F^{\alpha\beta} F_{\alpha\beta}$. 
A simple calculation shows that the combined two terms in (10) $S_{\lambda\mu\nu} + M_{\lambda\mu\nu}$ lead to the standard experession of the total angular momentum making use of ${\bf \nabla}.{\bf E}=0$, whereas the 
total expression for $J_{\alpha\mu\nu}$ directly gives this expression
\begin{equation}
{\bf L}_{EM} = \int {\bf r} \times ({\bf E} \times {\bf B}) d^3 x
\end{equation}
Note that the divergence term, i. e. the last term in expression (10) contains the factor
\begin{equation}
a_{\lambda\mu} = x_\lambda A_\mu -A_\lambda x_\mu
\end{equation}
Comparing it with the second term in (2) we find that $a_{\lambda\mu}$ is just the 4-dimensional version of ${\bf r} \times {\bf A}$. What is the significance of this? Further, the divergence term is usually made to vanish in the
integrals employing boundary conditions at spatial infinity. Could they have measurable consequences? We have proposed that geometric phase in optics has physical origin in terms of the angular momentum transfer caused by such a
divergence term \cite{23} ; van Enk and Nienhuis \cite{24} mention some evidence for this.

Proceeding with the similar mathematical manipulation for gluon field in QCD using the Lagrangian density
\begin{equation}
L_{QCD} = -\frac{1}{4} G^a_{\mu\nu} G^{a,\mu\nu}
\end{equation}
finally gives the angular momentum tensor $J^{QCD}_{\alpha\mu\nu}$ in which the divergence term reads
\begin{equation}
\partial^\sigma (x_\lambda A^a_\mu G^a_{\nu\sigma} - x_\mu A^a_\lambda G^a_{\nu\sigma})
\end{equation}
Here
\begin{equation}
G^a_{\mu\nu} = \partial_\mu A^a_\nu - \partial_\nu A^a_\mu - g f_{abc} A^b_\mu A^c_\nu
\end{equation}
Lowdon \cite{25} shows that the assumption to drop the boundary terms in the angular momentum decomposition is not justified and needs critical evaluation. Note that the last term in expression (3.12) of \cite{25} is exactly the
divergence term (14).

For the sake of completeness it may be pointed out that Corson's monograph \cite{22} contains a nice discussion on general gauge invariance (local gauge invariance for interacting matter-field system) in Sections 21 and 22, and derives
modified conservation laws using the Lagrangian density that depends on the gauge covariant derivative $D^\mu$ rather than the ordinary derivative $\partial^\mu$ of the matter field variables. Recently Lorce \cite{26} has used this
approach: pure gauge covariant derivative based on the Chen et al splitting of the gauge potential is defined to derive Noether currents and the field equations in QCD.

{\bf Relativistic invariance and Lorentz covariance}

Special theory of relativity can be discussed solely based on the algebraic approach without making use of the spacetime geometry. This was how Einstein in 1905 proved the relativistic invariance of the Maxwell-Lorentz electrodynamics;
Poincare and Lorentz had demonstrated the form invariance of the Maxwell equations under Lorentz transformations before the advent of Einstein's special relativity. Recall the chain of reasoning in proving the relativistic invariance of
the Maxwell equations by Einstein in \cite{27}. Maxwell equations represent experimental laws. Assuming charge density, $\rho$ to be a scalar the current density being a product of $\rho$ and velocity has to be a vector quantity in
three dimensional space. The form of Maxwell equations shows that electric field ${\bf E}$ is a vector but then the magnetic field ${\bf B}$ has to be an antisymmetric tensor. Lorentz transformation of various quantities leads to the
covariance of the Maxwell equations. Four dimensional spacetime geometric approach is more elegant. Generalized antisymmetric tensor $F_{\mu\nu}$ in 4-dimensions has only six independent components which could be identfied with 
${\bf E}, {\bf B}$, and assuming current density to be a 4-vector the Maxwell equations assume a manifest Lorentz covariant form; see Einstein's discussion \cite{27}.

Since EM potentials are not present in field equations the issue of gauge invariance does not arise. However in the Lagrangian formulation of the variational principle 4-vector EM potential $A_\mu$  is a fundamental dynamical variable.
Postulating 4-vector $A_\mu$ the EM field tensor $F_{\mu\nu}$ and the Lagrangian density are defined in the usual way. Gauge transformation (4) generalizes to
\begin{equation}
A_\mu ~ \rightarrow ~ A_\mu + \partial_\mu \chi
\end{equation}
The EM field tensor and the action are invariant under the gauge transformation (16). The question is whether this new symmetry, i. e. gauge symmetry affects the 4-vector character of $A_\mu$. It has been argued in \cite{9} 
that $A_\mu$ is not a true Lorentz 4-vector. Authors make two points: there is no consistent proof that $A_\mu$ is a 4-vector, and the standard textbook argument, e. g. in \cite{28} is just an argumentum in circulo. Authors 
introduce $A_\mu$ as a convenient auxiliary variable in the Maxwell equations, thus the inhomogeneous equation becomes
\begin{equation}
\partial_\mu \partial^\mu A^\nu -\partial^\nu (\partial_\mu A^\mu) =J^\nu
\end{equation}
Assuming Lorentz condition
\begin{equation}
\partial_\mu A^\mu =0
\end{equation}
Eq.(17) reduces to
\begin{equation}
\partial_\mu \partial^\mu A^\nu = J^\nu
\end{equation}
Since $J^\mu$ is a Lorentz 4-vector, and $\partial_\mu \partial^\mu$ is Lorentz invariant operator $A^\nu$ transforms as a Lorentz 4-vector, see p. 555 \cite{28}. On the face of it, there seems to be a cirularity as noted 
in \cite{9}. However one could directly focus on Eq.(17) and infer that $A^\mu$ is a 4-vector: since $\partial^\mu \partial_\mu$ is Lorentz invariant, $\partial^\nu$ is a Lorentz 4-vector and $J^\nu$ transforms as Lorentz 4-vector
$A^\nu$ has to transform as a Lorentz 4-vector. This argument is analogous to Einstein's reasoning to conclude that ${\bf B}$ is an antisymmetric tensor as mentioned above. The problem with (18) is that zero on rhs can have any Lorentz
transformation law depending on the lhs, therefore only if one assumes $A^\mu$ to be a 4-vector lhs is a Lorentz scalar \cite{9}.

It must be also realized that in the Lagrangian formulation $A_\mu$ is postulated to be a Lorentz 4-vector. Even if we ask for the proof of  the transformation of $F^{\mu\nu}$.
\begin{equation}
F^{\mu\nu}~ \rightarrow ~ \Lambda^\mu_\alpha \Lambda^\nu_\beta F^{\alpha\beta}
\end{equation}
it is obvious that the cited Einstein's arguments are adequate that rely on experiments. On the other hand, the transformation law for $A^\mu$ in \cite{9} citing earlier literature \cite{29,30,31} is
\begin{equation}
A^\mu ~ \rightarrow ~ \Lambda^\mu_\nu (A^\nu  + \partial^\nu \omega_\Lambda)
\end{equation}
Here $\omega_\Lambda$ is a Lorentz scalar, and the subscript $\Lambda$ associates it with the Lorentz transformation. Transformation (21) is consistent with (20) and the definition of $F_{\mu\nu}$, therefore, $A_\mu$ transforms as a
Lorentz 4-vector up to a gauge transformation. Since there is no unique $\omega_\Lambda$, there is an infinite number of physically equivalent Lorentz transformation laws. Let us examine the arguments carefully: $\omega_\Lambda$ is a
scalar function, $\partial^\nu \omega_\Lambda$ is obviously a Lorentz 4-vector, then the structure of (21) implies that $A^\nu$ must be a Lorentz 4-vector. How could arbitrariness in $A^\mu$ due to gauge symmetry lead to infinity of the
equivalent Lorentz transformation laws? There is only one Lorentz transformation law for $A^\mu$
\begin{equation}
A_\mu ~ \rightarrow ~ \Lambda^\nu_\mu A_\nu
\end{equation}
and the infinity of equivalent transformations (21) underline the fact that the arbitrariness in $A^\mu$ due to gauge transformations is superfluous.

Note that the internal gauge symmetry does not act on the spacetime, i. e. on the fundamental indefinite metric form $g_{\mu\nu} x^\mu x^\nu$ of the Lorentzian manifold, whereas Lorentz transformations are defined by
\begin{equation}
x^\mu ~ \rightarrow ~ \Lambda^\mu_\nu x^\nu
\end{equation}
such that the fundamental form remains invariant. The gauge group should not be mixed with the Lorentz group. Major controversial views in the proton spin debate have origin in this.

Moriyasu \cite{31} refers to Weyl's original gauge theory to introduce the modern gauge theories, and the review \cite{9} draws attention to the Christoffel symblos in general relativity to justify the generalized Lorentz 
transformation law (20). Levi-Civita's parallel displacement of a vector defines an affine manifold in Weyl's interpretation where the Christoffel symbol is viewed as an affine connection \cite{32}. Weyl's geometry has a spacetime
coordinate system and a gauge defined respectively by a quadratic differential form
\begin{equation}
dQ = g_{\mu\nu} dx^\mu dx^\nu
\end{equation}
and a linear form
\begin{equation}
d\phi = \phi_\mu dx^\mu
\end{equation}
General coordinate invariance is generalized to include gauge invariance; if the gauge is changed then
\begin{equation}
dQ~\rightarrow ~\lambda dQ
\end{equation}
and simultaneously
\begin{equation}
\phi_\mu ~ \rightarrow ~ \phi_\mu - \partial_\mu(log \lambda)
\end{equation}
Thus the gauge group is a noncompact group of homothetic transformations. Weyl identified the linear connection $\phi_\mu$ with the EM potentials $A_\mu$ in his unified theory.

Mathematically one could modify Weyl's idea such that $\phi_\mu$ defines a connection in a U(1) circle bundle over a Lorentzian spacetime such that the metric is unaffected by the gauge transformation. In this version, physical
interpretation of U(1) bundle as internal quantum phase leads to electrodynamics as a gauge theory. However the Lorentz transformations are determined by the coordinate transformations, and using the Lorentz tensors
to establish the Lorentz covariance of physical laws is not meaningless. Authors in \cite{9} assert that one may choose $\omega_\Lambda=0$ in (21) as a natural gauge and deal with the usual tensors. To make the Lorentz 
transformations contigent upon the choice of a gauge does not appear to be an advancement over the conventional relativistic invariance unless one has a wider group of transformations on the spacetime a la Weyl.

{\bf Gauge invariant extension}

In the proton spin controversy the idea of gauge invariant extension (GIE) evokes varied reactions \cite{9}. To appreciate its physical content the known but not sufficiently stressed elementary
aspects of gauge symmetry are recalled. For free field the gauge symmetry is contained in the invariance of the Lagrangian density under the gauge transformation defined by (16) that generalizes to the gluon field as
\begin{equation}
A^a_\mu ~ \rightarrow ~ U(A_\mu^a + \frac{i}{g} \partial_\mu)U^{-1}
\end{equation}
The extent of gauge transformations is unlimited and the number of independent dynamical variables, i. e. four for $A_\mu$, is more than the physically required. A subsidiary condition, e. g. the Lorentz condition (18) serves
two purposes: it limits the extent of gauge transformations and reduces the number of dynamical variables. Under the gauge transformation (16) the subsidiary condition (18) becomes
\begin{equation}
\partial_\mu A^\mu + \partial_\mu \partial^\mu \chi =0
\end{equation}
The Lorentz condition is manifestly Lorentz covariant but it is not gauge invariant unless
\begin{equation}
\partial_\mu \partial^\mu \chi =0
\end{equation}
Thus the arbitrary gauge transformation (16) is restricted to only those $\chi$ that satisfy (30). The number of independent dynamical variables is reduced to three; however there is a freedom to choose them from amongst
$(A_0, A_1, A_2, A_3)$.

In the Coulomb (or radiation) gauge the restriction is
\begin{equation}
{\bf \nabla}.{\bf A} =0
\end{equation}
This is not manifestly Lorentz covariant, and the gauge transformation (16) shows that for the gauge invariance to hold
\begin{equation}
\nabla^2 \chi =0
\end{equation}

In radiation field theory one is familiar with radiation electric and magnetic field vectors ${\bf E}^r, {\bf B}^r$ that satisfy the transversality condition
\begin{equation}
{\bf \nabla}.{\bf E}^r = {\bf \nabla}.{\bf B}^r =0
\end{equation}
Any 3-vector could in general be decomposed into transverse and longitudinal components. Thus the space component of 4-vector $A_\mu$ could be split into
\begin{equation}
{\bf A} = {\bf A}_t +{\bf A}_l
\end{equation}
and the transversality condition reads
\begin{equation}
{\bf \nabla}.{\bf A}_t =0
\end{equation}
Note that Eq.(35) is not equivalent to the Coulomb gauge condition: Wakamatsu has repeatedly stressed this point, e. g. in \cite{8,33}. Unfortunately he has mixed up the separation (34) with gauge symmetry inspired separation in
\cite{7}. We will return to this point in the next section. Substituting (34) and (35) in (31) we get
\begin{equation}
{\bf \nabla}.{\bf A}_l =0
\end{equation}
Longitudinal part is irrotational by definition but due to (36) it is also divergenceless.

The inhomogeneous Maxwell equations (17) written in terms of $A^\mu$ reduce to the form (19) in the Lorentz gauge. In the absence of sources both vector potential and scalar potential satisfy the homogeneous wave equation. In the
Coulomb gauge, the scalar potential satisfies the the Poisson equation and the vector potential satisfies the inhomogeneous wave equation that contains the gradient of the scalar potential. 
Jackson \cite{28} discusses this in Section 6.3.  

In QED it is usual to gauge transform the scalar potential by a suitable $\frac{\partial \chi}{\partial t}$ to set it to zero thereby reducing the independent variables to two satisfying the homogeneous wave
equation. Could it be related with hidden or mysterious Stueckelberg symmetry? Stoilov \cite{34} noted the significance of the Stueckelberg symmetry in the context of \cite{7} and Lorce has emphasized its crucial role in the uniqueness
issue and the path dependence of the gauge potential \cite{9, 35}. Nonuniqueness is also understood in terms of GIE \cite{33,36}. According to Lorce the Stueckelberg transformation on (34) is
\begin{equation}
{\bf A}_t ~ \rightarrow ~ {\bf A}_t +{\bf \nabla}C
\end{equation}
\begin{equation}
{\bf A}_l ~ \rightarrow ~ {\bf A}_l -{\bf \nabla}C
\end{equation}
where C is an arbitrary function. Wakamatsu argues that in QED the transverse-longitudinal decomposition is unique if a frane of reference is specified and C satisfies 
\begin{equation}
\nabla ^2 C=0
\end{equation}
Note that C may be time dependent, and $A_0, C, \chi$ obey the Laplace equation. Could the time derivative induce a nonzero scalar potential and result into the interplay between Stueckelberg transformation and gauge transformation?
Though this question has not been discussed in the literature, the authors in \cite{37} term the Stueckelberg symmetry as the original gauge symmetry in disguise.

Lorce has devoted considerable effort to interpret the Stueckelberg symmetry \cite{35}. He treats the Coulomb constraint in \cite{7} or light-cone (LC) constraint in \cite{38} as Stueckelberg-fixing constraints, and the 
Stueckelberg dependence as the background dependence. It is further clarified \cite{9} that strong and weak gauge invariances are distinguished by both Stueckelberg and gauge invariance in the former and only gauge invariance 
in the later. Thus Stueckelberg noninvariant GIE is weakly gauge invariant. On Wakamatsu's objections \cite{33} it is claimed that path dependence is not gauge dependence but the Stueckelberg dependence. that is even more general
than path dependence. It seems that in spite of some sort of agreement on the controversy there do exist differences \cite{11,12}.

The review \cite{9} nicely explains various facets of GIE, and also draws attention to its similarity with Lorentz invariant extension (LIE). Both GIE and LIE appear somewhat intriguing; interestingly there exists a concise discussion
on GIE and LIE on page 449 of \cite{39}. The rest mass is LIE of the energy from the rest frame: in the rest frame $(E, {\bf p}) =(m, 0)$, and Lorentz boosted $(E, {\bf p})$ shows that $m^2 =E^2 -p^2$. This is the standard relation in
any inertial frame of reference. Note that if one has massless particle there is a problem with LIE. The transverse component ${\bf A}_t$ is GIE of ${\bf A}$ in the Coulomb gauge. The Coulomb gauge (31) to generic gauge frame
is defined by the gauge transformation (16) such that
\begin{equation}
\chi = -\frac{1}{\nabla^2} {\bf \nabla}.{\bf A}
\end{equation}
gives the gauge invariant transverse vector potential in the generic gauge frame
\begin{equation}
 {\bf A}_t ={\bf A} -\frac{1}{\nabla^2} {\bf \nabla} ({\bf \nabla}.{\bf A})
\end{equation}
It is easily recognized that (41) can be re-written in terms of the projection operators \cite{33}
\begin{equation}
P^{ij}_T = \delta^{ij} -\frac{\nabla^i \nabla^j}{\nabla^2}
\end{equation}
Wakamatsu rightly asserts on the uniqueness of the decomposition (34) under the stated assumptions \cite{33}. Leader and Lorce \cite{9} give a detailed discussion on GIE bringing out some important points. The traditional 
wisdom says that a gauge noninvariant quantity is not measurable, however one could seek a gauge invariant quantity and fix a gauge such that it is identical with the gauge noninvariant quantity - thus GIE is measurable. Though
Jaffe-Manohar decomposition is gauge noninvariant, its LC-GIE is measurable that corresponds to Hatta decomposition \cite{38}. The role of nonlocal operation (e. g. the inverse Laplace operator in (41)) is crucial. It is possible to
discuss nonlocality in terms of Wilson lines. Wilson line in LC-GIE runs along a path defined by $n^\mu =\frac{1}{\sqrt 2} (1, 0, 0, -1)$ and reduces to unity in the LC gauge constraint
\begin{equation}
A^+ =0
\end{equation}
Choice of a path is suggested to be related with the Stueckelberg symmetry. Curiously Coulomb-GIE is path independent, cannot be understood in terms of Wilson lines and is nonlocal.

Above discussion shows that fundamental concepts have been re-analyzed from different angles inspired by the controversy that followed \cite{7}. Leader has reminded us that in QED what matters is the matrix elements
of the operators, therefore, state vectors in the Hilbert space and the role of gauge invariance of the operators have to be carefully understood \cite{18}. Physical observables and measurable quantities ultimately decide 
the validity of a theory; unfortunately in QCD one must deal with two regimes - perturbative QCD and strong coupling. Besides even in QED measurement is an intricate issue, for example, Schwinger \cite{40} points out that 
'microscopic measurement has no meaning apart from a theory', and further that, 'measurements individually associated with different regions in space-like relation are causally independent, or compatible'.
It is also fairly established that experimentally demonstrated Aharonov-Bohm effect is a nonlocal phenomenon. Thus in spite of better understanding on the controversial issues related with the proton spin problem there
still remain diverse physical interpretations. Possibly it is due to two pathways for the theory: generalization and reduction. For example, if one begins with the Coulomb gauge and seeks generalization for its validity
in every inertial frame then together with the Lorentz transformation a suitable gauge transformation is also required. In both GIE and LIE a sort of generalization is implied. However there could be unavoidable
arbitrariness in the process. Trautman gives an example \cite{41} how a noncovariant equation could be transformed into a covariant form introducing arbitrary new auxiliary fields. The noncovariant equation
\begin{equation}
A_1 =0
\end{equation} 
assumes general covariant form
\begin{equation}
u^\mu A_\mu =0
\end{equation}
where $u^\mu$ is the coordinate basis vector field.

On the other hand, in a reduction process it may happen that some important feature gets lost. The assumed vanishing of total divergence terms, for example, in the last term in Eq.(10) and the expression (14). The best option
would be to minimise the arbitrariness and check the self-consistency.

\section{\bf Gauge invariant spin decomposition}

The objective of gauge invariant decomposition of QCD total angular momentum into quark and gluon angular momenta led Ji \cite{5} to use symmetric and gauge invariant Belinfante energy-momentum tensor. The resulting angular momentum
density can be decomposed into quark and gluon parts. Chen et al \cite{7} recall that Ji decomposition, though gauge invariant, does not separate total gluon angular momentum into spin (SAM) and orbital (OAM) parts, and that quark 
OAM and gluon angular momentum operators do not satisfy the rotation group algebra
\begin{equation}
[J_i , J_j] = i \epsilon_{ijk} J_k
\end{equation}
Naturally their claim to have solved this 'long-standing' problem of gauge invariant QCD angular momentum decomposition aroused great interest in the QCD community. They first gave the example of QED to illustrate the new idea.

The starting point of their work is the standard QCD canonical angular momentum
\begin{align}   
J_{QCD} &=\int [\psi^\dagger \frac{1}{2}{\bf \Sigma} \psi +\psi^\dagger {\bf x} \times \frac{1}{i} {\bf \nabla} \psi  \nonumber \\
     &\qquad +{\bf E}^a \times {\bf A}^a +E^{ai} {\bf x} \times {\bf \nabla} A^{ai}] d^3x
\end{align}   
One often writes Eq.(47) as $J_{QCD}={\bf S}_q +{\bf L}_q +{\bf S}_g +{\bf L}_g$ in which except the first term, all other are nongauge invariant. The suggested solution is to make all the gauge noninvariant expressions
gauge invariant by making following replacements
\begin{equation}
{\bf A}^a ~ \rightarrow ~ {\bf A}^a_{phys}
\end{equation}
\begin{equation}
{\bf \nabla} ~ \rightarrow ~ {\bf D}_{pure} = {\bf \nabla} -i g {\bf A}^a_{pure} t^a
\end{equation}
The principal new idea is to decompose the gauge potential into a pure gauge part and a physical part; in the 4-vactor form it reads
\begin{equation}
A^a_\mu = A^{a, pure}_\mu +A^{a, phys}_\mu
\end{equation}
The gauge transformation (28) is given the following prescription
\begin{equation}
A^{a, pure}_\mu ~ \rightarrow ~ U[A^{a, pure}_\mu +\frac{i}{g} \partial_\mu] U^{-1}
\end{equation}
\begin{equation}
A^{a, phys}_\mu ~\rightarrow ~ U A^{a, phys}_\mu U^{-1}
\end{equation}
Evidently by construction physical part is assumed gauge covariant and the gauge symmetry of the gauge vector potential (28) is entirely contained in the pure part Eq. (51). The authors note that the actual
determination of pure and physical parts in (50) is a nontrivial task; however they suggest generalized curl-free condition for the pure part
\begin{equation}
{\bf D}^{pure} \times {\bf A}^{a, pure} t^a =0
\end{equation}
and for physical part the vanishing of the commutator
\begin{equation}
[{\bf A}^{a, phys} t^a , {\bf E}^a t^a] =0
\end{equation}
The condition (54) is obtained by imposing gauge invariance on the gluon angular momentum.

It is clear that the proposed solution crucially depends on the unambiguous determination of the pure and the physical parts in (50) otherwise the whole exercise of transforming (47) into the gauge invariant form with the 
replacements (48) and (49) amounts to an atrificialty. Unfortunately the condition (54) is already based on the circularity: the proposed split (50) leads to the gauge invariance of ${\bf L}_g$, but
the gauge invariance of ${\bf L}_g$ is imposed to define ${\bf A}^{a, phys}$.

Since the gauge symmetry is fundamental to motivate (50) we term Chen et al proposal as gauge symmetry paradigm (GSP). To investigate whether GSP gives any new physics two aspects have to be looked into: How is the theory 
modified beginning at the level of the action principle? What are the new mathematical and physical consequences of the basic postulate (50)? Since the whole effort in \cite{7} is focused on the rearrangement of $J_{QCD}$ 
terms it lacks a solid foundation. The first attempt to remedy this situation seems to be that of Zhou and Huang \cite{42}, and more recently by Lorce \cite{26}. They obtain the results on angular momentum decomposition
in conformity with \cite{7}, and also get the standard QED and QCD field equations. At first it seems surprising that no new result is obtained. One criticism \cite{17} of the derivation given by Lorce is that
the constraint on pure field tensor to be zero was used at the initial level of the construction of the Lagrangian density itself, and hence it lacks rigour. Now we find that nothing is altered even if we adopt the Lagrange 
multiplier procedure to incorporate the constraint. The reason is that the separation (50) is a trivial artefact with no new physics.

The QED example illustrates it in a transparent manner. For the EM field Lagrangian density the separation of $A_\mu$ into pure and physical parts gives
\begin{align}
L_{EM} &= -\frac{1}{4}F^{\mu\nu,pure}F_{\mu\nu,pure} +F^{\mu\nu,phys} F_{\mu\nu,phys}   \nonumber \\
  &\qquad +2 F^{\mu\nu,pure} F_{\mu\nu,phys}+C_{\mu\nu}F^{\mu\nu,pure}
\end{align}
The constraint
\begin{equation}
F^{\mu\nu,pure}=0
\end{equation}
is imposed using the Lagrange multiplier antisymmetric tensor $C_{\mu\nu}$. Both $A^{\mu,pure}$ and $A^{\mu,phys}$ are treated as independent dynamical variables in the variational process; two field equations are obtained that show
that for their consistency we get the Maxwell equation and
\begin{equation}
\partial^\mu C_{\mu\nu} =0
\end{equation}
and variation $\delta C_{\mu\nu}$ leads to (56). Note that the presence of $L_{Dirac}$ for the electron field and considering the total Lagrangian density for electron+Maxwell field does not alter the consequence that the standard
Maxwell-Dirac equations are obtained. The same conclusion holds for QCD.

The second question is very important as the entire theory is built on the proposition (50)-(52). In the light of new insights gained in the preceding section it becomes somewhat easier to address this question. We begin with the QED
example as presented in \cite{7}: their equations (6) to (10). A unique decomposition of the vector potential
\begin{equation}
{\bf A} = {\bf A}_{pure} +{\bf A}_{phys}
\end{equation}
with the gauge transformations
\begin{equation}
{\bf A}_{pure} ~ \rightarrow ~ {\bf A}_{pure} +{\bf \nabla} \chi
\end{equation}
\begin{equation}
{\bf A }_{phys} ~ \rightarrow ~ {\bf A}_{phys}
\end{equation}
define the new idea; note that Eqs. (58)-(60) are the special cases of Eqs. (50)-(52). A subtle assumption is made that pure and physical parts in (58) are determined by
\begin{equation}
{\bf \nabla}.{\bf A}_{phys} =0
\end{equation}
\begin{equation}
{\bf \nabla} \times {\bf A}_{pure} =0
\end{equation}
Authors state that, 'These are nothing but the transverse and longitudinal components of the vector potential ${\bf A}$'.

Several points are in order keeping in mind the critique in Sec.II.

(i) Transverse-longitudinal decomposition is based on the inertial frame of reference whereas pure-physical decomposition is based on the gauge transformations. The two are conceptually and physically different.

(ii) In the transverse-longitudinal decomposition (34) the number of the degrees of freedom remains three, but in (58) it is doubled. Moreover the Coulomb gauge condition further reduces the number of the independent
dynamical variables in the former. Thus the philosophy to increase the number of dynamical variables is contrary and inferior to the sound conventional approach aimed at reducing the redundance.

(iii) Conditions equivalent to (61) and (62) in the transverse-longitudinal paradigm (TLP) define the decomposition itself. In contrast, in the GSP (59) and (60) define the components; additional constraints (61) and (62)
or their reinterpretation in the language of gauge transformations renders the new idea vacuous.

(iv) The statement from \cite{7} quoted above is misleading reflecting the medley of relativity and gauge symmetry; most strikingly in the wider context of its Lorentz covariant generalization by Wakamatsu \cite{43}.

From the beginning the lack of the simultaneous manifest Lorentz covariance and gauge invariance was noticed \cite{13}, however in the present paper we arrive at a new result: the source of the controversy lies
in the flaw at the fundamental level of pure-physical decomposition identified with the transverse-longitudinal decomposition, i. e. the confusion between GSP and TLP. It is for this reason that there is a
continued disagreement on the uniqueness issue between Wakamatsu and Lorce. Wakamatsu rightly emphasizes that the transverse-longitudinal decomposition and the Coulomb constraint are not identical, and the transversality
condition ensures uniqueness of the decomposition based on the Helmholtz theorem \cite{33}. However his assertion that Chen et al decomposition is nothing new it is just transverse-longitudinal decomposition
is not correct. In U(1) gauge theory (QED) superficially it seems so, but the decomposition (58)-(60) is certainly not transverse-longitudinal decomposition. Lorce clearly follows the new idea, and what he terms
as the Stueckelberg symmetry makes sense only in this context \cite{35}. The decomposition (50) to (52) is arbitrary up to the transformation
\begin{equation}
A^{a,pure}_\mu ~ \rightarrow ~ A^{a,pure}_\mu +B^a_\mu
\end{equation}
\begin{equation}
A^{a,phys}_\mu ~ \rightarrow ~ A^{a,phys}_\mu - B^a_\mu
\end{equation}
\begin{equation}
B^a_\mu = \frac{i}{g} U_{pure} U^{-1}_S [\partial_\mu U_S] U^{-1}_{pure}
\end{equation}
\begin{equation}
U_{pure} ~ \rightarrow ~ U U_{pure}
\end{equation}
Here $U_S$ is the unitary Stueckelberg symmetry transformation matrix; it is gauge invariant. He notes that Stueckelberg symmetry is broken by the Coulomb constraint, and $U_{pure}$ or $A_{pure}$ is a
background field. Hence the conclusion that Chen et al decomposition is nonunique becomes unavoidable. Unfortunately seemingly similar transformations (39) and (40) are also viewed as Stueckelberg transformations both
by Lorce and Wakamatsu; our discussion shows they are not.

The textbook approach to modern gauge theories usually begins with the gauge invariance of complex scalar field or Dirac spinor field. For example, free electron Dirac Lagrangian density is shown to possess global 
gauge invariance, imposing local gauge invariance leads to the necessity of a gauge potential that couples to the Dirac field. Maxwell term or the kinetic energy term for the gauge potential is added to make
it a genuine dynamical field. However purely the requirement of local gauge invariance is insufficient to introduce $A_\mu$. It is easily seen that the Dirac Lagrangian density
\begin{equation}
L_{Dirac} =\bar{ \psi} (i\gamma^\mu \partial_\mu -m)\psi
\end{equation}
is invariant under the global gauge transformation, and for the local gauge transformation
\begin{equation}
\psi ~ \rightarrow ~ e^{i\alpha(x)} \psi
\end{equation}
one could add a term $(\partial_\mu \phi)\bar{\psi} \gamma^\mu \psi$ where $\phi ~ \rightarrow ~\phi+\alpha$ to render the total Lagrangian density gauge invariant. Note that the conserved Noether current $\bar{\psi} \gamma^\mu \psi$
makes the added term a pure divergence term $\partial_\mu(\phi \bar{\psi} \gamma^\mu \psi)$. Thus the local gauge invariance has no dynamical consequence. It is not surprising that Lorce finds that in QCD pure gauge
potential in Chen et al decomposition assumes the form
\begin{equation}
A^{a,pure}_\mu = \frac{i}{g} U_{pure} \partial_\mu U^{-1}_{pure}
\end{equation}
Thus the gauge symmetry inspired decomposition turns out to be trivial. In reality, we assume dynamical gauge field based on experimental laws embodied in Maxwell field equations such that 
$-\frac{1}{4} F^{\mu\nu}F_{\mu\nu}$ is gauge invariant under the gauge transformation $A_\mu ~ \rightarrow ~ A_\mu + \partial_\mu \alpha$, and then discover local gauge symmetry (68) for the interacting Dirac-Maxwell field system.

In a formal way postulating local gauge transformation (68) for Dirac spinor and gauging the symmetry defining gauge covariant derivative $D_\mu$ that transforms as
\begin{equation}
D_\mu \psi ~ \rightarrow ~ e^{i\alpha(x)} D_\mu \psi
\end{equation}
one obtains the standard QED Lagrangian density
\begin{equation}
L_{QED} =-\frac{1}{4} F^{\mu\nu}F_{\mu\nu} + \bar{\psi}(i \gamma^\mu D_\mu -m)\psi
\end{equation}

To summarize: the principal idea to split the gauge potential into pure and physical parts solely based on the gauge symmetry has not been developed to its logical conclusion in the literature. The application of constraints
and supplementary conditions makes the decomposition useful in practice but simultaneously the idea is metamorphosed into the conventional theory.

\section{\bf Topological Approach}

The main result of the last two sections is that Chen et al decomposition of the gauge potential examined from different perspectives finally proves to be vacuous. Therefore it is futile to continue the proton spin 
decomposition controversy and seek re-interpretations apparently reconciling the conflicting views \cite{6,11,12,14}. Note that the concepts of nonlocal operators and generalized Lorentz transformations did
not originate due to Chen et al proposal \cite{9}. The GIE-LIE issue is also not new \cite{39}. Leader and Lorce recognize the superfluousity in Chen et al approach due to 'two four-vector fields at one point', see the paragraph
following Equation (191) in \cite{9}. In the differential geometric approach \cite{37} limited to investigate only the gauge potential decomposition (50) it is concluded that nothing is gained from this decomposition.

Recent articles \cite{11,12} show that differences do persist: pDIS experiments select a preferred direction, however Lorce interprets this as 'actual experimental conditions determine the form of the background field', while 
according to Wakamatsu this implies 'the requirement of boost-invariance along the direction of the nucleon momentum'. More surprising is the confusion on the transverse and the physical components of the gauge
potential in \cite{6,14,15}. Ji had criticized \cite{44} the Chen et al proposal earlier, however re-emphasizing the role of infinite momentum frame and the natural physical gauge (LC gauge) the frame dependence of
the gluon spin operator is argued in \cite{6,14}. It is intriguing that the authors replace $A_\mu^{a,pure}$ and $A_\mu^{a,phys}$ with the longitudinal and transverse components of $A_\mu^a$ respectively without providing any
explanation. Note that Equation (5) in \cite{6} and Equation (14) in \cite{14} are not identical to the original Chen et al suggestion where one has instead $A_\mu^{a,pure}$ and $A_\mu^{a,phys}$. Curiously Wang et al
\cite{15} approvingly refer to the work of \cite{14} and use the term 'physical transverse part'. Since we have shown that frame dependent decomposition and gauge symmetry inspired decomposition are distinct the claims made in 
\cite{6,14,15} have to be viewed with serious reservations.

Setting aside the controversy, the current status of the proton spin problem reviewed in the literature \cite{2,8,9,10,11,45} indicates that the standard sum rules and the decomposition of the angular momentum operators
for calculating physical observables employing perturbative QCD have not succeeded in solving the problem. Could there be a crucial role of nonperturbative QCD in the proton spin problem? An interesting conjecture
put forward by Bass on a nonlocal gluon topological structure \cite{2} has not received the attention it deserves; one of the reasons could be the extreme intricacies in the nonperturbative QCD. Though not directly
related with this, Lowdon has made an important contribution examining the role of boundary terms in QCD \cite{25}. QCD angular momentum decomposition following Balinfante procedure is developed, and it is shown that the
standard method to drop divergence terms is unjustified. Its consequence on the proton spin sum rules is negative: their validity is doubtful. The expression for total angular momentum operator can be written as
\begin{equation}
J^i_{QCD} = L^i_q +S^i_q + L^i_g +S^i_g +(S^i_1+S^i_2)
\end{equation}
The last two terms on rhs represent the surface terms. Assuming them to vanish the matrix elements between z-polarized proton states yield the proton spin sum rule
\begin{equation}
\frac{1}{2} = \frac{1}{2}\Sigma +L_q +S_g+L_g
\end{equation}
Lowdon first shows that the surface terms do not necessarily vanish. Irrespective of it he further shows that the matrix elements of $S^i_1$ and $S^i_2$ cancel those for the quark spin $S^i_q$ and gluon spin $S^i_g$ operators
respectively. How does one interpret this remarkable result? Lowdon himself hints towards a nonperurbative QCD approach to spin problem. Even though Lowdon's discussion is based on Belinfante's procedure the importance
of boundary terms is not limited to this.

It is well established that perturbative QCD and factorization theorems play very important role in understanding DIS and pDIS experiments \cite{3}. However the structure functions embody the characteristic properties
of quasi-free quarks and gluons a la asymptotic freedom of QCD, and one may doubt proton spin in terms of the angular momenta of the partons just as one cannot obtain proton mass from the partons. In SU(3) isospin multiplets,
for example, $J^P=\frac{1}{2}^+$ octet and $J^P = \frac{3}{2}^+$ decuplet of baryons one qualitatively understands the mass spectrum of the baryons in terms of the symmetry breaking. In contrast, the scenario for spin is 
entirely different: the spin $\frac{1}{2}$ or $\frac{3}{2}$ is fixed for octet or decuplet of baryons irrespective of the nature of the constituent quarks. We suggest that spin has topological origin; may be it is related with 
the nontrivial QCD vacuum/monopoles. Unfortunately the understanding of QCD vacuum is still in the exploratory state, nevertheless a tentative topological idea for spin problem has been recently proposed \cite{16}. This
idea utilizes the geometric concept of Wilson loops and de Rham theorems. Further progress in this approach is reported here.

Wilson lines or gauge links have been recently discussed in the context of proton spin problem \cite{33,46,47}, and a nice introduction is also given in \cite{3}. In the topological approach Wilson loops
and nonabelian Stokes theorem are fundamental geometric objects, and de Rham periods to understand topological obstructions, for example, monopoles require extreme care in the nonabelian gauge theories. Aharonov-Bohm
phase and Dirac monopoles motivate a physical picture, however more technical details are necessary to develop the proposition on de Rham theorems in QCD made in \cite{16}. Following Fishbane et al \cite{48} and
Marsh \cite{49} we present the essentials.

As explained in \cite{16} the language of differential forms is natural in this approach. Differential 1-form of the 4-vector gauge potential is denoted by
\begin{equation}
A=A^a_\mu t_a dx^\mu
\end{equation}
In \cite{49} the notation used is $A_\mu = gA^a_\mu t_a$. A continuous curve parameterized by s defines a path starting at $x_0$ and ending at x by $z^\mu (s=1) =x^\mu$ and $z^\mu (s=0) =x^\mu_0$, and a Wilson
line appears in the solution of parallel transport equation
\begin{equation}
W(x, x_0) = P exp[ig~\int^x_{x_0} ~\frac{dz^\mu(s)}{ds} A_\mu(z(s)) ds]
\end{equation}
Geometrically (74) is interpreted as connection 1-form. Since the path is divided into an infinite number of infinitesimal elements the matrix valued integrals have to preserve a path ordering P. Under a gauge
transformation Wilson line transforms as
\begin{equation}
W(x, y) ~\rightarrow ~ U(x) W(x, y)U^{-1}(y)
\end{equation}
The combination $\bar{\psi}(x)W(x, y) \psi (y)$ is gauge invariant; it is for this reason that a Wilson line is inserted between quark field operators at different positions. This gauge invariant combination depends
on the end points $(x, y)$ as well as the path that connects them. Of particular interest in QCD is the light-front coordinate description defined by $x^\pm =(t\pm z)/\sqrt 2$ and $x^\mu=(x^+, x^-, {\bf x}_t)$.
For light-like separation the Wilson line along $x^-$-axis depends only on the end points \cite{3}, and one can replace a direct line $W(x, y)$ by $W(x, +\infty)W(+\infty, y)$.

A wilson loop is defined by a closed path such that $z_\mu(0)=z_\mu(1)$
\begin{equation}
W(L) = Tr P exp[ig ~ \oint A]
\end{equation}
Stokes theorem relates the line integral with the surface integral over the surface enclosed by the loop. The nonabelian Stokes theorem \cite{48, 49} is stated to be
\begin{equation}
P exp[ig\oint_{\partial S} A] = P_S exp[ig\int_S W(a, y) G(y) W(y, a)]
\end{equation}
The curvature 2-form is defined as
\begin{equation}
G = \frac{1}{2} G_{\mu\nu} dx^\mu \wedge dx^\nu =dA +A\wedge A
\end{equation}
On the rhs of (78) the symbol $P_S$ denotes surface ordering of infinitesimal area elements \cite{48}, $\partial S$ is the boundary of surface S, and a is a fixed reference point on the boundary. Introducing the notations
\begin{equation}
\bar{A}(y) =W(a, y) A(y) W(y, a)
\end{equation}
\begin{equation}
\bar{G}(y) = W(a, y) G(y) W(y, a)
\end{equation}
and using the properties of $W(a, y)$ the Poincare lemma is obtained
\begin{equation}
d\bar{A} =\bar{G}
\end{equation}
\begin{equation}
d\bar{G} =dd\bar{A} =0
\end{equation}
Here $\bar{G}$ and $d\bar{G}$ are gauge invariant apart from a position-independent gauge transformation at the reference point U(a), and (82) is gauge invariant and also independent of the reference point a.
In the derivation of (83) use has been made of the Bianchi identity \cite{49}.

If there is no nontrivial topological structure, specially the absence of monopoles, then the nonabelian Stokes theorem is sufficient to define de Rham periods. However the interpretation of surface integral as a flux
is not unique since the field tensor $G_{\mu\nu}$ itself carries color. The presence of monopoles introduces additional complications: it has been shown that the QCD flux through a closed surface can be defined in the
loop space parameterized by two parameters s and t \cite{49}. The variation of parameter t from 0 to $2\pi$ gives a closed 2-dimensional space-like syrface, and the closed surface $\Sigma$ obtained through loop
space replaces S on the rhs of (78) together with the change of G to $\bar{G}$
\begin{equation}
\Theta (\Sigma) = P_tP_s exp [ig \oint_\Sigma \bar{G}]
\end{equation}
We have used symbolically the path orderings $P_tP_s$ as the Wilson line (75) appearing in the integral on the rhs of (84) has to be defined for each value of the loop space parameter t
\begin{equation}
W_{z_t} (s,0) =P_s exp[ig \int^s_0 A_\mu(z_t(s)) \frac{dz^\mu_t(s)}{ds} ds]
\end{equation}

The integral of a p-form over a cycle defines a period of the form where cycle is analogous to a loop integral \cite{16}. Marsh points out that at least for 2-forms one has unambiguous generalization of de Rham
theorems to the nonabelian case. Note that his first and second de Rham theorems correspond to the second and first theorems respectively in our paper \cite{16}. The first de Rham theorem states that for a given set of periods
$[\nu_i]$ there exists a closed 2-form $\omega$ such that
\begin{equation}
\nu_i = \oint_{C_i} \omega
\end{equation}
$[C_i]$ is a set of independent 2-cycles. The second theorem states that the closed 2-form $\omega$ is exact if all its periods give the identity matrix. It is in contrast to the abelian case in which all periods of an exact 
form vanish \cite{16}.

In the QCD the 2-form in (86) is $\bar{G}$, and the implication of second theorem is that a unique $\bar{A}$ does not exist. This arises because of the arbitrariness of the gauge rotation with position independent
U(a) on the gauge potential and the field tensor mentioned earlier. Wu and Yang \cite{50} as well as Fishbane et al \cite{48} have argued that nonabelian flux through a loop cannot be defined, however Marsh has clarified
that the arbitrariness due to path dependence is limited to position independent gauge transformations, and the nonabelian flux is a meaningful quantity. We discuss it below for the proton spin problem.

In an important paper Burkardt \cite{51} explained the difference between Ji and Jaffe-Manohar definitions of quark OAM, and also gave a physical interpretation to the potential angular momentum suggested by
Wakamatsu \cite{33}. Recently Burkardt has updated his work \cite{45}; see also \cite{11}. In the present context we are interested only in the geometry of paths and Wilson lines. The ingenuity in Burkardt's
work lies in a U-shaped light-like path. In the light front coordinates $(0^-, {\bf 0}_t)$ is linked to $(\infty^-, {\bf0}_t)$ then to $(\infty^-,{\bf x}_t)$  and the path is completed returning to $(x^-,{\bf x}_t)$.
The wilson line to achieve gauge invariance in the Wigner distributions is constructed of three straight line gauge links by Burkardt
\begin{align}
W_U^{+LC} &= W(0^-, {\bf 0}_t;\infty^-,{\bf 0}_t)~ W(\infty^-,{\bf 0}_t; \infty^-,{\bf x}_t) ~ \nonumber \\
     &\qquad W(\infty^-,{\bf x}_t; x^-, {\bf x}_t)
\end{align}.
In the LC gauge the gauge potential at light-cone infinity is pure gauge potential, and assumed to satisfy antisymmetric boundary conditions
\begin{equation}
{\bf A}_t(+\infty,{\bf x}_t) = -{\bf A}_t(-\infty, {\bf x}_t)
\end{equation}
Burkardt's aim is to calculate transverse momentum and OAM using two different paths: a direct straight line path from $(0^-,{\bf 0}_t)$ to $(x^-,{\bf x}_t)$, and a staple to $\pm \infty$, i. e. the U-shaped
path described above. From PT invariance it is argued that quark OAM calculated from $W^{+LC}_U$ and $W^{-LC}_U$ is equal, and the Jaffe-Manohar OAM can be taken equal to either of them. On the other hand Ji OAM
is obtained by a straight line path having 
\begin{equation}
W_{straight} = W(0^-,{\bf 0}_t; x^-, {\bf x}_t)
\end{equation}
The difference between the two is identified with the potential OAM.

We are interested in a closed path to define a Wilson loop. It is straightforward to see that a U-shaped path followed by a straight line path from $(x^-,{\bf x}_t)$ to $(0^-,{\bf 0}_t)$ results
into the desired closed path. The associated Wilson loop is given by
\begin{equation}
W_{QCD}(L) = W^{+LC}_U ~ W(x^-,{\bf x}_t: 0^-,{\bf 0}_t)
\end{equation}
Stokes theorem (84) shows that (90) leads to the nonabelian color flux enclosed by the loop , and the gauge field is given by
\begin{equation}
\bar{G} (x^-,{\bf x}_t) = W(0^-,{\bf 0}_t; x^-, {\bf x}_t) G(x^-,{\bf x}_t) W(x^-,{\bf x}_t; 0^-,{\bf 0}_t)
\end{equation}
Expression (91) offers the color flux interpretation of the potential OAM consistent with Burkardt's interpretation. However the additional new aspect is that of the possibility of a quantized flux in the presence of 
topological obstructions.

Intuitively the color field-free interior of the proton having nontrivial topological objects like monopoles could be envisaged. Unfortunately monopoles in the nonabelian gauge theories not only have different definitions but 
also have intricate mathematical properties: Wu-Yang monopoles and 't Hooft-Polyakov monopoles have been extensively discussed in the literature. Chan et al \cite{52} and Marsh \cite{49} use loop space formalism
to discuss Wu-Yang monopole. Kondo's monopole \cite{53} is both interesting and intriguing for several reasons. First, it is akin to 't Hooft-Polyakov magnetic monopole but obtained in pure Yang-Mills theory 
without Higgs scalar field. Second, the Diakonov-Petrov version of the nonabelian Stokes theorem is used. And finally the gauge potential is decomposed into two parts
\begin{equation}
A^a_\mu = V^a_\mu + X^a_\mu
\end{equation}
in which $V^a_\mu$ transforms like the original gauge potential $A^a_\mu$ under the gauge transformation, and $X^a_\mu$ transforms like an adjoint matter field; color confinement is due to the former, i. e. $V^a_\mu$.

There is a kind of universality in the assignment of spin for the members of each of the baryon multiplets. Does it indicate a topological invariance for the spin? It would be interesting to explore such a possibility.
Since the angular momentum is a third rank tensor a 3-form would represent it; a natural choice is
\begin{equation}
\bar{S} = \bar{A}~ \wedge ~\bar{G}
\end{equation}
and formally the corresponding de Rham period can be defined over a 3-cycle (3-C)
\begin{equation}
\oint_{3-C} \bar{S} = topological ~invariant
\end{equation}
In the abelian gauge theory expressions (93) and (94) are easily understood. Consider periods over 1-cycle and 2-cycle of 1-form A and 2-form F respectively, then using the Kuenneth product rule
\begin{equation}
\oint_{3-C} A \wedge F =\oint_{1-C} A ~ \oint_{2-C} F
\end{equation}
if $F=dA$ the period (95) vanishes for $1-C =\partial (2-C)$. There is an interesting case when ${\bf E}.{\bf B} \neq 0$ then the period (95) is nonzero. For QCD the generalization of de Rham theorems 
for 3-forms and the Kuenneth rules are not obvious. However this problem and the role of flux quantization and monopoles in the proton spin problem deserve attention to unravel QCD physics. At present we have not been
able to make a definitive contribution in this direction.

\section{\bf Conclusion}
A thorough study on the proton spin decomposition controversy is reported with the important result that the gauge symmetry inspired decomposition of the gauge potential proposed by Chen et al has not been
taken to its logical development in the literature: either it becomes vacuous or it is metamorphosed into the conventional transverse-longitudinal based decomposition. It is argued that the proton
spin problem going beyond the controversy and the spin sum rules requires a nonperturbative QCD approach. Topological ideas are suggested to understand the spin of baryons - some kind of topological invariants. Nontrivial 
QCD vacuum structure and mathematical complications in defining de Rham theorems for the nonabelian gauge theories have to be overcome for further advances in this approach. A curious similarity between Chen et al 
decomposition (50) and Kondo decomposition (92) indicates that a consistent development of the Chen et al proposal in the spirit of topological considerations discussed by Kondo may lead to interesting new physics in 
the QCD spin problem.

{\bf Acknowledgment}

I gratefully acknowledge useful correspondence on the proton spin problem with M. Wakamatsu, C. Lorce, and E. Leader.

{\bf APPENDIX}

There seems to be a lack of appreciation on the applicability and the uses of de Rham theorems in QCD. The main aim of this appendix is to bring out the fact that de Rham cohomology is not restricted to the Riemannian manifolds, and has natural extension to the nonabelian gauge fields.

Cartan exterior differential forms and the integration over chains in a given manifold M of dimension n formally define the Stokes theorem
\begin{equation}
\int_c d\omega =\int_{\partial c} \omega
\end{equation}
where the domain of integration c has the boundary $\partial c$, and $d\omega$ is the exterior differential of the differential form $\omega$. The formal structure of (96) has the profound inplication: the topology of the integration domains (homology) and the topology of the differential forms (cohomology) are dual to each other. Metric-independence of (96) ensures its applicability to a Lorentzian spacetime. To make it explicit let us consider the definition of a p-form. A totally anti-symmetric covariant p-tensor field using local coordinates and the exterior or wedge product defines the p-form $\omega$
\begin{equation}
\omega =\omega_{i_1 ....i_p}~ dx^{i_1}\wedge dx^{i_2} ...dx^{i_p}
\end{equation}
The Hodge star operator defines a unique dual (n-p)-form
\begin{equation}
^*\omega = \frac{\sqrt g}{{(n-p)}!} \epsilon^{i_1...i_p}_{i_{p+1}...i_n} ~dx^{i_{p+1}} \wedge ...dx^{i_n} 
\end{equation}
Here g is the determinant of the metric tensor of the Riemannian manifold. The volume n-form $\tau$ defines the Hodge dual such that for every $\alpha$ belonging to the space of p-forms $\Lambda^p(M)$ the inner product is
\begin{equation}
\tau (\alpha|\omega) =\alpha \wedge *\omega
\end{equation}
For a Lorentzian manifold the appropriate sign changes are needed in the definitions (97) and (98): the determinant is $\sqrt{-g}$ and the Hodge star is
\begin{equation}
*^2 =-1{(-1)}^{p(n-p)}
\end{equation}

The homology of M is the set of equivalence classes of cycles $Z_p$ of degree p that differ by boundaries $B_p$ i. e. $H_p(M) =Z_p/B_p$. On the other hand, de Rham cohomology $H^p(M)$ is defined as the equivalence set of closed differential forms $Z^p$ modulo the set of exact forms $B^p$ i. e. $H^p =Z^p/B^p$. The dimension of $H_p(H^p)$ is called the Betti number $b_p(b^p)$ of M. The global properties of a manifold are best understood using de Rham theorem that could be stated as the duality of the vector spaces $H_p$ and $H^p$, and $b_p=b^p$ for finite dimensional $H_p$ and $H^p$; it also establishes the duality of $H^p$ with compact support to the infinite chain homology for orientable manifolds \cite{54}.

An alternative form of this theorem is obtained defining a de Rham period. The integral of a closed form over a cycle is called a period. Obviously de Rham period depends on the homology class of the cycle and the cohomology class of the differential form. For the topological photon model \cite{55} we used the first and the second de Rham theorems as follows:

{\bf {First de Rham Theorem}} For a given set of periods $[\nu_i]$ there exists a closed p-form such that
\begin{equation}
\nu_i =\oint_{c_i} \omega
\end{equation}
where $[c_i]$ is a set of independent cycles. The closed form is determined up to the addition of an exact form.

{\bf {Second de Rham Theorem}} If all the periods of a closed form vanish then it is exact.

Note that the second theorem stated as above is given as corollary in \cite{54}.

Finally there are two important technical issues in generalizing the de Rham theorems to the nonabelian gauge theories. First the symmetry gauge group SU(N) necessitates a Lie algebra valued matrix differential form. Secondly, the nonabelian gauge field equations being nonlinear the harmonic p-form defined by the Laplacian becomes complicated.

\end{document}